\documentclass[12pt]{article}
\makeindex

\usepackage{amsmath,amssymb,array,calc,rotating,epsfig,psfrag}
\usepackage{graphicx}
\usepackage{hyperref}
\usepackage{mathtools,slashed}
\usepackage[utf8]{inputenc} 
\usepackage[dvipsnames]{xcolor}
\usepackage{tensind}
\usepackage{cancel}
\usepackage{tensor}
\usepackage{cleveref}
\usepackage{multirow}
\tensordelimiter{?}
\usepackage[font=scriptsize,labelfont=bf]{caption}
\def\a{\alpha}
\def\b{\beta}

\def\O{\Omega}

\def\o{\omega}
\def\m{\mu}
\def\g{\gamma}
\def\l{\lambda}
\def\n{\nu}

\def\t{\tau}
\def\s{\sigma}
\def\a{\alpha}
\def\b{\beta}
\def\d{\delta}
\def\g{\gamma}
\def\G{\Gamma}

\def\rd{{\rm d}}

\def\tJ{{\tilde{J}}}

\def\cD{{\mathcal{D}}}

\def\cK{{\mathcal{K}}}

\def\tkappa{{\tilde{\kappa}}}
\def\tJ{{\tilde{J}}}

\def\be{\begin{equation}} 
\def\ee{\end{equation}}

\begin{document}
\begin{titlepage}
\centerline{\bf \Large  From Virasoro Algebra to Cosmology } 

\vskip 1.5cm 
\centerline{\bf \footnotesize   Vincent G. J. Rodgers\footnote{vincent-rodgers@uiowa.edu} }

\vskip 1cm
\centerline{ Department of Physics and Astronomy}
\centerline{ \it  The University of Iowa}
\centerline{\it Iowa City, IA 52242}

\date{\today}

\begin{abstract}
Earlier work of Balachandran and friends provided a map from algebras to field theories.  These methods  provide insight into quantum gauge theories and anomalies.  In this note we take the reader from the coadjoint representation of the Virasoro algebra to four- (and higher-) dimensional gravitation and cosmology. The protagonist in this story is a component of the projective connection, the diffeomorphism field,  which  straddles between the one-dimensional world of  initial data  in string theories to cosmology in four dimensions.  We review mathematical intuition that ties projective geometry to the Virasoro algebra,   the   
Thomas\textendash Whitehead (TW) gravitational action that gives the diffeomorphism field dynamics and  the building blocks for gauge projective Dirac action.
\\

\noindent To be published in \textit{Particles, Fields, and Topology: Celebrating A.P. Balachandran,} a Festschrift volume for A.P. Balachandran (World Scientific Publishing Co., Singapore) \end{abstract}
\end{titlepage}

\section{Introduction}
While a post-doc at Stony Brook in 1986, I would often venture back to Syracuse and listen in at the original Room 316 discussions directed by Professor Balachandran.  As always, the Room 316 meetings were intellectually rich and inundated with ideas and perhaps far too many ideas than one could humanly thrash out.  On one occasion,  Professor Balachandran introduced the idea of geometric actions that arise on coadjoint orbits of Lie groups.  He and friends \cite{Balachandran:1983oit,Zaccaria:1981fi,Balachandran:1986hv,Balachandran:1987st} had been able to show that starting from an algebra and its dual,  one could construct geometric actions that correspond to  core components  in certain quantum field theories.  One of his students, Balram Rai, and I took up Bal's challenge to build the geometric actions from that Kac\textendash Moody and Virasoro algebras \cite{Rai:1989js}\;.  We  recovered the Wess\textendash Zumino\textendash Witten model from the Kac\textendash Moody algebra and the two-dimensional Polyakov action from the Virasoro algebra.  This work was later extended to the super-Virasoro algebra \cite{Delius:1990pt}\;.  I remember running to Balachandran, out of breath, to tell him how, not only did his method of coadjoint orbits give the anomalous contribution to effective actions of these two-dimensional theories but it also got the couplings correct.      I did not expect this. But Prof. Balachandran's heart  barely skipped a beat and he said, with the coolness of Hercule Poirot, ``Don't be so surprised. That's what I was expecting.'' 

\begin{figure}[ht]
\renewcommand*\figurename{2D Polyakov Action}
\begin{equation}
S = \frac{\tilde c}{2 \pi} \int d x d \tau \left[
\frac{\partial^2_{x} s}{(\partial_{x}s)^2} \partial
_{\tau} \partial_{x} s - \frac{(\partial^2_{x}s)^2
(\partial_{\tau} s)}{(\partial_{x} s)^3} \right]
 - \int d x d \tau  \,\,{D}(x)
\frac{(\partial s/ \partial \tau)}{(\partial s/ \partial
x)}. \label{2DPolyakov} 
\end{equation}
\caption{The 2D Polyakov action that arises using the method of coadjoint orbits. The geometric action relates the coadjoint element $\mathcal{D}=({D},\tilde c)$ to the field ${D}$ and the coupling constant $\tilde c$ in the action.   ${D}$, called the \emph{diffeomorphism field}, is the protagonist in this origin story.}
\label{eq:Polyakov}
\end{figure}

To sketch how this works, let us consider the Virasoro algebra \cite{OvsienkoValentin2005Pdgo,Segal:1981ap,Witten:1987ty}  as the centrally extended algebra of vector fields in one dimension.  Let $(\xi,a)$ and \(({\eta},b)\) be two such vectors with respective central elements,  $a $  and $b$. The commutator
extends to,\be 
[({\xi}, a), ({\eta},b) ] = ({\xi \circ \eta}, ((\xi,\eta))_{0}),\; 
\ee
where $\xi \circ \eta$ is defined by 
\begin{equation}
\xi \circ \eta \equiv \xi^a \partial_a \eta^b - \eta^a \partial_a \xi^b,
\end{equation} and $((\xi,\eta))_{0} $, the so-called Gelfand\textendash Fuchs two-cocycle \cite{Gelfand}\;, is defined by 
\begin{align}
\label{Gelfand-Fuchs1} 
((\xi,\eta))_{0} &\equiv
 \frac{c}{2\pi} \int (\xi \eta''' )\,d\theta - (\xi \leftrightarrow \eta)\\ &=\frac{c}{2\pi} \int \xi^a\nabla_a (\boldsymbol{[g^{bc}\nabla_b \nabla_c ]}\eta^m)g_{mn}d\theta^n - (\xi \leftrightarrow \eta)\;,  
\label{Gelfand-Fuchs2} 
\end{align}and where $g_{ab}$ represents  one-dimensional metric.  Eqs.~(\ref{Gelfand-Fuchs1}) and (\ref{Gelfand-Fuchs2}) form an invariant pairing between $\xi$ and $\eta'''$.  In the Gelfand\textendash Fuchs two-cocycle,  we see that   the    vector field $\eta$  has been mapped into a \emph{quadratic differential}, $\mathcal{Q}_{an}(\eta)$, by the insertion of a Laplacian, $\boldsymbol{ g^{bc}\nabla_b \nabla_a}$, into the  Lie derivative. Explicitly, using abstract index notation,\begin{equation}
\eta \partial_\theta  \rightarrow \eta''' d\theta^2= \nabla_a ([ g^{bc}\nabla_b \nabla_c] \eta^m)g_{mn}d\theta^a d\theta^n = \mathcal{Q}_{an}(\eta)d\theta^a d\theta^n.\label{mapping}
\end{equation} The Gelfand\textendash Fuchs two-cocycle is the simplest  example of an invariant pairing between a vector and a quadratic differential, $D$, that is also centrally extended by the element $\tilde c$, viz., $\mathcal{D}=(D, \tilde c)$.  
\be  < (\xi,a) | (D,\tilde c) > \equiv \int (\xi D) d \theta+  a \tilde c= \int (\xi^i D_{ij}) d \theta^j +a \tilde c\;. \label{pair} \ee  The  invariance of this  pairing requires that      
\be
ad^*_{(\eta,d)}(D, \tilde c)=(\eta D' + 2 \eta' D -  \tilde c\, \eta''',0),\;
 \label{coadjointaction}
\ee
defining the coadjoint representation of the Virasoro algebra \cite{Kirillov:1982kav,Witten:1987ty}\,. Now,  Kirillov  observed \cite{Kirillov:1982kav}  a symplectic two-form on the coadjoint orbits of the Virasoro algebra. For each coadjoint element, $\mathcal{D}=(D, \tilde c)$ and a pair of coadjoint elements $\mathcal{D}_1$ and $\mathcal{D}_2$ that are infinitesimally close to $\mathcal{D}$ through  $ ad^*_{(\xi,b)}(D, \tilde c)$ and $ad^*_{(\eta,a)}(D, \tilde c) $ respectively, one has the two-form, $\Omega_\mathcal{D}(\mathcal{D}_1, \mathcal{D}_2),$ on the  orbit of $\mathcal{D}$ given by 
\begin{equation}
\Omega_\mathcal{D}(\mathcal{D}_1, \mathcal{D}_2) = \frac{ \tilde c}{2\pi} \int (\xi \eta''' - \xi'''
  \eta)\,dx
+ \frac{1}{2\pi} \int (\xi \eta' - \xi' \eta )D\,dx\;.
\label{2cocycle2}
\end{equation}
 $\Omega$ is a symplectic two-form as its anti-symmetric, closed and non-degenerate. This observation   \cite{Rai:1989js} is what led to Eq.(\ref{2DPolyakov}).    The method of coadjoint orbits has dissolved the coadjoint element $\mathcal{D}$  into a coupling constant and a component of a two-dimensional field $D$,  called the \emph{diffeomorphism field},  in two dimensions.   In Eq.(1), it appears as a background field coupled to the Polyakov metric, $
\frac{(\partial s/ \partial \tau)}{(\partial s/ \partial
x)}$, and is often interpreted as an anomalous energy\textendash momentum tensor.  A similar situation occurs in the Kac\textendash Moody case where the coadjoint element there, $\mathcal{A}=(A, \hat b),$ allows us to interpret $A$ as the remaining component of a  two-dimension gauge-fixed Yang\textendash Mills connection $A_\mu$.  It corresponds to a background Yang\textendash Mills field that is coupled to a Wess\textendash Zumino current.  We seek a similar geometric connection interpretation  of $D$ that, like Yang\textendash Mills, can be represented in any dimension.     In the spirit of a connection, one also sees that the Gelfand\textendash Fuchs case, Eq.(\ref{Gelfand-Fuchs1}), lives in the ``{pure gauge}'' sector since it corresponds to the  $\mathcal{D}=(0,c)$ coadjoint element. 

An important clue as to the geometric natures of $D$ comes from, yet another, observation by Kirillov.  He observed \cite{Kirillov:1982kav} that Eq.(\ref{coadjointaction}) provides a correspondence  to the  space of Sturm\textendash Liouville operators.   Thus there is a  one-to-one correspondence,
\be 
(D, \tilde c) \Leftrightarrow -2 \tilde c \frac{d^2}{dx^2} + D(x),
\label{correspondence}\ee
between the coadjoint element  $(D, c)$ and a Sturm\textendash Liouville operator with weight $\tilde c$ and Sturm\textendash Liouville potential,  $D(x)$. This  important observation ties the Virasoro algebra into projective geometry.  Let's explain. 
 
To see this correspondence \cite{Tabachnikov,OvsienkoValentin2005Pdgo}\;, let
 \(\phi_A\) and  \(\phi_B\) be the two independent solutions  of the Sturm-Liouville equation that span the space of solution,   
\begin{equation}
 (-2 \tilde c \frac{d^2}{dx^2} + D(x)) \phi_A=0, \hskip.3in
 (-2 \tilde c \frac{d^2}{dx^2} + D(x)) \phi_B=0. \label{SturmLiouville2}
\end{equation}
 Now define the ratio, \(f(x)= \frac{\phi_A(x)}{\phi_B(x)}\).  Then,  one can show that, 
\begin{equation}
D(x) = \frac{ \tilde c}{2} \left( \frac{f'''(x)}{f^{'}(x)}- \frac{3}{2} \left(\frac{f''(x)}{f^{'}(x)}\right)^2\right)=S(f(x)).
\end{equation} This is precisely the projective  invariant, the Schwarzian derivative of \(f(x)\)  with respect to \(x\) \cite{Tabachnikov,OvsienkoValentin2005Pdgo}\;,    \(S(f(x))\). Therefore we may consider \(f(x) \) to be an affine parameter, \(\tau \equiv f(x)\), on the projective line \(\mathbb{P}^1\) of   a one-parameter family of Sturm\textendash Liouville operators, viz.
 \begin{displaymath}
L_\tau \phi  = -2 \tilde c \frac{d^2}{dx^2} \phi + D_\tau \phi =0. \label{tauSturm}
\end{displaymath}
   This above equation is invariant under the action of the vector field  $(\eta, d) $, and in particular, the Sturm\textendash Liouville operator transformation, $ad^*_{(\eta,d)}\,L_\tau  $, is as in Eq.(\ref{coadjointaction} ). 
  This begins the identification of ${D }$ as a \emph{projective connection}\cite{Kirillov:1982kav} and to be interpreted \cite{Brensinger:2017gtb} as a component of the Thomas-Whitehead connection, $\tilde \nabla_\mu$ \cite{Thomas:1925a,Thomas:1925b,Whitehead}\;. We discuss the salient features of this presently and circle back to this identification.

\section{Thomas\textendash Whitehead and  Projective Curvature}
\subsection{The TW Projective Connection}
The simplest  way to appreciate projective curvature is through the parameterization of geodesics and connections that yield the same geodesics.    If  $\hat \nabla$ and $\nabla$ admit the same geodesics, they belong to the same projective equivalence class. Thomas showed how one can write a gauge theory over this projective symmetry \cite{Thomas:1925a,Thomas:1925b}\;.

    Consider a $\rd$-dimensional manifold $\mathcal{M}$ with coordinates $x^a$ where italic latin indices $a,b,c,m,n,\dots = 0,1,\dots, \rd-1$.  
Let  $\hat{\nabla}_a$ be a connection on $\mathcal{M}$ where  \(\zeta^a \) is geodetic, i.e.
\begin{align}
        \zeta^b \hat{\nabla}_b \zeta^a =\frac{d^2 x^a}{d\tau^2} + \hat\Gamma^a{}_{bc} \frac{d x^b}{d\tau}\frac{d x^c}{d\tau}= 0. & \label{geo1}
\end{align}
Now consider another connection related to the previous connection by a one-form $v_a$.  This is called a \emph{projective transformation}, 
\begin{equation}
\Gamma^a{}_{bc} = \hat \Gamma^a{}_{bc} + \delta^a_{\,\,\,b} v_c + \delta^a_{\,\,\,c} v_b. \label{ProjectiveTransformation}
\end{equation}
The geodesic equation for this connection is then 
\begin{align}
        \zeta^b {\nabla}_b \zeta^a =\frac{d^2 x^a}{d\tau^2} + \Gamma^a{}_{bc} \frac{d x^b}{d\tau}\frac{d x^c}{d\tau} = f(\tau)\frac{d x^a}{d\tau},  & \label{geo2}
\end{align}
where \(f(\tau) = 2 v_b \frac{d x^b}{d\tau}.   \)
We can eliminate the  right-hand side by suitable reparameterization of  $\tau$ to some  $u(\tau)$.  Because of this,  both Eqs.~(\ref{geo1}) and (\ref{geo2}) admit the same geodesic curves.   The connections belong to the same projective equivalence class, $\hat \Gamma^a{}_{bc} \sim  \Gamma^a{}_{bc} $.   

Thomas \cite{Thomas:1925a,Thomas:1925b}\;  developed a ``gauge" theory of projectively equivalent connections in the following way. First,  define the   \emph{fundamental projective invariant} $\Pi^{a}{}_{bc}$ 
\begin{align}\label{e:Pi}
        \Pi^{a}{}_{bc} \equiv \Gamma^a{}_{bc} - \tfrac{1}{(\rd+1)} \delta^a_{\,\,(b}{} \Gamma^m{}_{c)m}.
\end{align}
This is clearly invariant under a projective transformation, Eq.(\ref{ProjectiveTransformation}). The geodetic equation associated with $\Pi^{a}{}_{bc}$ is, 
\begin{align}
        \frac{d^2 x^a}{d\tau^2} + \Pi^a{}_{bc} \frac{d x^b}{d\tau}\frac{dx^c}{d\tau}=0,
\end{align}
is projectively invariant but not covariant. This is because  $\Pi^{a}{}_{bc} $ does not transform as a connection under a general coordinate transformation from \(x \rightarrow x'(x) \) due to the last summand in 
\be
{\Pi'}^a_{\ bc} = J^a_{\ f} \left( \Pi^f_{\ de} \bar{J}^d_{\ b} \bar{J}^e_{\ c} + \frac{\partial^2 x^f}{\partial x'^b \ \partial x'^c} \right)
 + \frac{1}{d+1} \frac{\partial \log |J|}{\partial x^d} \left( \bar{J}^d_{\ b}\delta^a_{\ c} + \bar{J}^d_{\ c}\delta^a_{\ b} \right).  \label{e:PiTrans}\ee
Here \(J^a_{\ b} = \frac{\partial x'^a}{\partial x^b}\), the Jacobian of the transformation and 
  \(J=\det{(J^a_{\ b})} \). Thomas  constructs a line bundle (also called volume bundle) over $\mathcal{M}$ to form a  $\rd +1$-dimensional manifold,  $\mathcal{N,}$ referred to as the Thomas Cone \cite{Eastwood}\;. The coordinates on the Thomas Cone are  \(( x^0, x^1, \dots , x^{\rd -1} , \lambda  )    \), where $\l$ is a fiber denoting  the volume coordinate.  It is because the  volume coordinate, $\l,$ takes values from $0< \l < \infty$, that  $\mathcal N$ is called a cone. These coordinates transform as 
\begin{align}\label{e:ConeTrans}
        x'^\a = ( x'^0(x^d), x'^1(x^d), \dots , x'^{\rd -1}(x^d) , \lambda' = \lambda |J|^{-\frac{1}{\rm{d}+1}} ),
\end{align}
and one can see this is a fibration as the $\l$ coordinate is not independent. For every coordinate transformation on  $\mathcal{M}$ there is a unique coordinate transformation on  $\mathcal{N}$. We use Greek indices   over coordinates on  $\mathcal{N}$  and italic Latin indices   over coordinates on $\mathcal{M}$.  We reserve the index $\lambda$ and the upright letter $\rd$ to refer to the volume coordinate  $x^\rd = x^\lambda =\lambda$. 

To proceed, let $\Upsilon$ denote the fundamental vector  on $\mathcal{N}$ and a companion one-form $\omega_\a$ such that $\Upsilon^\a \o_\a =1$. For any function on $\mathcal{N}$, $\Upsilon^\a \nabla_\a f = \l \partial_\l f. $     The projective  connection, $ {\tilde \nabla}_\a $,  must be compatible with  $\Upsilon,$ meaning that the covariant derivative on $\Upsilon$  returns the identity. Thus $\Upsilon$ satisfies the normalized geodesic equation, viz.,   
\be {\tilde \nabla}_\a \Upsilon^\b = \d_\a^{\;\b} \; \Rightarrow \; \Upsilon^\a{\tilde \nabla}_\a \Upsilon^\b= \Upsilon^\b.\ee
Explicitly, we can take $\Upsilon^{\alpha} = (0,0,\dots,\lambda)$,
 so we can  write a generalized connection ${{\tilde\G}}^{\a}_{\,\,\b \g}$ \cite{Crampin,Whitehead}\;on $\mathcal{N}$ as, 
\begin{equation} 
{{\tilde\G}}^{\a}_{\,\,\b \g}= \begin{cases}
    {\tilde \G}^{\lambda}_{\,\,\,\lambda a}={\tilde \G}^{\lambda}_{\,\,\, a \lambda} = 0
    \\ {\tilde \G}^{\a}_{\,\,\,\,\lambda \lambda} = 0 \label{e:Gammatilde}\\ {\tilde \G}^{a}_{\,\,\,\,\lambda b}={\tilde \G}^{a}_{\,\,\,\,b \lambda} = \o_\l\d^a_{\,\,b}\;\;\\
{\tilde \G}^{a}_{\,\,\,\,b c} ={ \Pi}^{a}_{\,\,\,\,b c}\\
{\tilde \G}^{\lambda}_{\,\,\,\, a b} =  \Upsilon^{\l} \cD_{ a b}\;
 \end{cases} 
\end{equation} 
We have only required  $\mathcal{D}_{ab}$ to transforms  on $\mathcal{M}$ as,
\be
        \cD'_{ab} = \frac{\partial x^m}{\partial x'^a}\frac{\partial x^n}{\partial x'^b} (\cD_{mn} -\partial_m j_n-j_m j_n+j_c\Pi^c{}_{mn}),\label{DiffTransformation} 
\ee
  so that the connection $\tilde \G^{\m}_{\,\,\,\a \b}$   transforms as an affine connection
on $\mathcal{N}$\cite{Brensinger:2020gcv}\;. 
Here,   $j_a = \partial_a \log{|J|^{-\frac{1}{\rm{d}+1}}}$.
 The covariant derivative operator now transforms covariantly on the Thomas cone,  i.e. $\nabla'_\a = \frac{\partial x^\b}{\partial x'^\a} \nabla_\b$.  
In the above, $\cD_{ab}$ generalizes the work of Thomas and is  independent of \({ \Pi}^{a}_{\,\,\,\,b c}\). This is the  origin of the diffeomorphism field \(\mathcal{D}\). 
\vskip.2in
\begin{scriptsize} 
\textbf{
First Relation to Virasoro Algebra:} In one dimension, $\mathcal{D}_{ab}$ transforms in one-to-one correspondence with the coadjoint element $\cD$\cite{SamBrensinger}\;. Indeed, in one dimension, Eq.(\ref{DiffTransformation}) collapses to
\( \delta_\xi \mathcal{D}_{11} = 2\xi'\mathcal{D}_{11} + \mathcal{D}_{11}'\xi - \frac{1}{2}\xi''',   \) for a vector field $\xi$.
Now let \(\mathcal{D}_{11} = qD\) where \( q=\frac{1}{2 \tilde c}.\) Then    \( \delta_\xi (qD) \Rightarrow \delta_\xi D = 2\xi'D + D'\xi - \frac{1}{2q}\xi'''. \) This is as in Eq.(\ref{coadjointaction}).
\end{scriptsize}
\vskip.2in
Furthermore, a vector field  $\chi$ on $\mathcal{M}$ may be promoted to a vector field $\tilde \chi$ on $\mathcal{N}$ by writing \(\tilde \chi^\a = (\chi^a, -\l x^b \kappa_b),\) where $\kappa_a $ only needs to transform as 
\begin{equation}
\kappa'_a = \frac{\partial x^m}{\partial x'^a} \kappa_m - \frac{1}{\rd+1} \frac{\partial \log J}{\partial x'^a}.
\end{equation} 
  Similarly, a one-form  $v$ on $\mathcal M$ can be related to a projective one-form $\tilde v$ via  
 \( \tilde v_\b = (v_b+ \kappa_b, \frac{1}{\l}).\) Thus,  any metric $g_{ab}$ on $\mathcal{M,}$ accompanied with its \(g_a \equiv -\frac{1}{\rm{d}+1}\,\partial_a \log \sqrt{|g|},\)  can be promoted to a metric $G_{\a \b}$ on $\mathcal{N }$ as, \begin{align}\label{e:BigGSuccinct}
        G_{\a\b} &= \delta^a_{\,\,\alpha} \delta^b_{\,\,\beta} \,g_{ab} - \lambda_0^2 g_\alpha g_\b\\
        G^{\a\b} &= g^{ab} (\delta^\alpha_{\,\,a} - g_a \Upsilon^\a)(\delta^\b_{\,\,b} - g_b \Upsilon^\b) - \lambda_0^{-2} \Upsilon^\a\Upsilon^\b,
\end{align} 
with $g_\a \equiv (g_a, \frac{1}{\l})$ and $\l_0$ a constant.
\vskip.2in
\begin{scriptsize} \textbf{Second Relation to Virasoro Algebra:} The projective  two-cocycle, defined  by (inserting the projective Laplacian as in Eq.(\ref{mapping})) the two-cocycle:\[
<\xi,\eta>_{(\zeta)} =\tilde{c} \int_{C(\zeta)} \xi^\a ({\tilde \nabla}_\a \boldsymbol{[G^{\rho \n} {\tilde \nabla}_\rho {\tilde \nabla}_\n ]}\eta^\b\ G_{\b \m} )\zeta^\m d\s-(\xi \leftrightarrow \eta),
\]   for a path $C$ parameterized by $\s$.    The vector, $\zeta^\mu =(\zeta^b,-\l \zeta^a g_a )$ defines the path $C$. This collapses to the Kirillov two-cocycle, Eq.~(\ref{2cocycle2}), for paths restricted to $\mathcal{M}$, $\zeta_{0}^\mu =(\zeta^b,0)$. 
\[
<\xi,\eta>_{(\zeta_{0})} \ =\tilde{c} \int \xi_1 \left(2 \mathcal{D}_{11}- g_{11} \frac{1}{\l_0^2 } \right)\eta'_1 dx 
+\tilde{c} \int \xi_1 \eta_1^{'''}dx -(\xi \leftrightarrow \eta)\;.
\]
 Comparing this to Eq.(\ref{2cocycle2}), we make the observation that the projective connection and the coadjoint element $(D, \tilde{c})$ are in correspondence through
\[
2\tilde{c}\,  \mathcal{D}_{11}-  \frac{\tilde{c}}{\l_0^2 } g_{11}=  D,
\] where the projective connection is shifted by the one-dimensional metric tensor $g_{11}$.   
 \end{scriptsize}
 \vskip.2in
\subsection{Geodesics revisited}
For insight, let us revisit the    geodesics on $\mathcal{M. }$   Consider a  geodetics on $\mathcal{N}$, \begin{equation}
\zeta^\a \tilde \nabla_\a \zeta^\b =0 \;.
\end{equation}
Separating the $\mathcal{M}$ coordinates from the fiber $\l$, we have the expressions
\begin{eqnarray}
&&\frac{d^2 x^a}{du^2}+ { \Pi}^{a}_{\,\,\,\,b c}\frac{dx^b}{du}\frac{dx^c}{du} = -2 \frac{1}{\l} \left(\frac{d\l}{du}\right)\frac{dx^a}{du}, \label{Aeq} \\
&& \frac{d^2 \l}{du^2} + \l \cD_{bc} \frac{dx^b}{du}\frac{dx^c}{du}=0.\label{Beq}
\end{eqnarray} 
Together, these equations are covariant and projectively invariant.  Now reparameterize    $u$ to a  parameter $\t$ that is affine with respect to the projective invariant ${ \Pi}^{a}_{\,\,\,\,b c} $ in  Eq.(\ref{Aeq}).  Then with  $u \rightarrow \tau(u),$ we find that the   Schwarzian derivative of  $\t$ with respect to $u$, $S(\t (u))$, must satisfy,   
\begin{equation}
 \cD_{bc} \frac{dx^b}{du}\frac{dx^c}{du} = \frac{1}{2} \frac{\frac{d\t}{du} (\frac{d^3 \t}{du^3})- \frac{3}{2} (\frac{d^2 \t}{du^2})^2}{ (\frac{d\t}{du})^2}\equiv \frac{1}{2} S(\t(u))\;. \label{Schwartzian Derivative}
\end{equation}  
As an  example, if  $\cD_{bc} \frac{dx^b}{du}\frac{dx^c}{du}$ vanishes, then $\t = \frac{a u + b}{c u +d} $, where $a,b,c,$ and $d$ are real numbers.  This is the  M\"obius  transformation. Another  example is when   $\cD_{bc} \frac{dx^b}{du}\frac{dx^c}{du} = \frac{m^2-1}{2 u^2} $. Then   $\t = (\frac{a u^m + b}{c u^m +d})$ is the requirement.           M\"obius transformations are  one-dimensional projective transformations. \(\Pi^a_{\ bc}\) and \(\mathcal{D}_{bc}\) are the  components of  a \textit{projective connection},  $\tilde \nabla_\b$.  From here we can compute curvature,  spin connections, as well as the Dirac operator on the Thomas cone. 
\section{Building Blocks for  TW Gravity}
Equipped with the connection ${\tilde \nabla}_\a$, we can compute  the projective curvature tensor, \( 
[{\tilde \nabla}_\a,{\tilde \nabla}_\b] V^\g = \,{\cK}^{\g}_{\,\,\,\rho\a \b  } V^\rho. 
\)
The only non-vanishing components are\[ \cK^{a}_{\ bcd} = \mathcal{R}^{a}_{\ bcd} + \delta^a_{\ [c}\mathcal{D}_{d]b} \;\;\text{and}\;\; \cK^{\lambda}_{\ cab} = \lambda \partial_{[a}\mathcal{D}_{b]c} + \lambda \Pi^{d}_{\ c[b}\mathcal{D}_{a]d},   \]
where $\mathcal{R}^{a}_{\ bcd} $ is the projective equivariant curvature\cite{Thomas:1925a} ``tensor'' that is constructed using $\Pi^a_{\;\; b c}$ instead of $\Gamma^a_{\;\; b c}$ in the Riemann curvature tensor.  
The \emph{Thomas\textendash Whitehead action}\cite{Brensinger:2017gtb}  can be constructed as:
\( S_\text{TW} = S_{\text{PEH}} + S_{\text{PGB}}, \)
where the projective Einstein\textendash Hilbert action is 
\begin{equation} \label{eq:PEH Action} S_{\text{PEH}} = -\tfrac{1}{2\tkappa_0 \lambda_0} \int d\lambda\ d^{\rm{d}}x \sqrt{|G|} \cK^{a}_{\ bcd}(\delta^{c}_{\ a} g^{bd}) \end{equation}
and the projective Gauss-Bonnet action is
\begin{equation} \label{eq:PGB Action} S_{\text{PGB}} =- \tfrac{\tJ_0 c}{\lambda_0} \int d\lambda \ d^{\rm{d}}x \sqrt{|G|} \ \left(\cK^{\alpha}_{\ \beta \gamma \rho}\cK_{\alpha}^{\ \beta \gamma \rho} - 4\cK_{\alpha \beta}\cK^{\alpha \beta} + \cK^2\right). 
\end{equation}
The fields \( \cD_{ab}\), \(\Pi^f_{\ de} \) and \( g_{ab}\) are independent fields.  This action will give dynamics to \( \cD_{ab} \) in both two and  four-dimensions and  collapses to the Einstein-Hilbert action plus a topological term when  \(\Pi^f_{\ de} \)  is compatible with   \( g_{ab}\) and when \( \cD_{ab} =0.\)  This is a Palatini\cite{palatini} formalism of the action.    
\vskip.2in \begin{scriptsize} \textbf{Third Relation to Virasoro Algebra:}
The  interaction term in the two-dimensional Polyakov action arises from the projective Einstein\textendash Hilbert.  Let $g_{ab}$ denote the covariant Polyakov metric.  Then the interaction term in the two-dimensional Polyakov action becomes,
\begin{align} \label{eq:TwoD Projective EH Action} \begin{split} S_{\text{PEH}} &= -\tfrac{1}{2\tkappa_0 \lambda_0}\int d^2x \; d\l \sqrt{|G|} \cK_{\a\b} G^{\a \b}    = -\tfrac{1}{2\tkappa_0 \lambda_0}\int d^2x\; d\lambda \sqrt{|G|}   \cK   \\ &= \frac{1}{{\kappa}_0} \int d x d \tau  \,\,{\mathcal{D}}(x)
\frac{(\partial s/ \partial \tau)}{(\partial s/ \partial
x)}=\frac{1}{{\kappa}_0} \int d x_+ d x_{-}   \;\cD_{++} h_{--}\;, \end{split} \end{align} and $\kappa_0$ absorbs an integration of $\l$ over specific limits. The last equality expresses the coupling in light-cone coordinates\cite{Rai:1989js}\;.\end{scriptsize}
\vskip.2in 
We can continue to add building blocks and promote fermions to the Thomas cone.      The gamma matrices on (even-dimensional) $\mathcal{M}$ easily extend to  $\tilde\g^\a$, consistent with the anti-commutation relations that give the metric, one finds that 
 \(\tilde{\gamma}^{m} = \gamma^{m}\) and \(  \tilde{\gamma}^{\lambda} = - \frac{\lambda}{\lambda_0} \left( i \gamma^5 + \lambda_0g_{m}\gamma^{m} \right) ,\) where $\gamma^5$ is as usual on $\mathcal{M}$.
We can also build the spin connection using the frame fields of the metric,  \( \tilde\omega^{\mu}_{\ \a \nu} =  \partial_{\nu}\tilde e^{\mu}_{\ \a} + \tilde \Gamma^{\mu}_{\ \rho \nu} \tilde e^{\rho}_{\ \a} \;.   \)
It contains the diffeomorphism field in its $\l$ component and will lead to a pseudoscalar coupling to the fermions\cite{Brensinger:2020gcv}\;,
\[\tilde{\omega}_{\underline 5\underline b m} = -\lambda_0 e^{
p}_{\ \underline b} \left( \mathcal{D}_{p m} - \partial_{m}g_{p} + \Pi^{n}_{\ p m} g_{n}  \right). \]
With this we can define the covariant derivative acting on the spinors   as
\[\tilde \nabla_{\mu} = \partial_{\mu} + \tilde \O_{\mu}\; \text{    where \;      }
 \tilde \O_{\mu} = \frac{1}{4}\tilde \omega_{\underline \a \underline \b \mu} \tilde\gamma^{\underline \a} \tilde\gamma^{\underline \b} .\]
The fermions, themselves can  promoted  to the  Thomas cone    
  \cite{Brensinger:2020gcv} through a simple $\l$-dependent matrix-valued phase with $v$ and $w$ corresponding to  density and chiral density weights,  \[\Psi(x^a,\l)= \left(\frac{\l}{\l_0}\right)^{\frac{(\rm{d}+1)}{2}(v I_{4}+ w \g^5)}\phi(x^a). \]

\section{Results and Discussion}
  The  coadjoint orbit methods, developed by Balachandran and collaborators provided the initial insight  into the  meaning of the Virasoro algebra in   higher-dimensional gravities.  The  correspondence between the Virasoro algebra and projective geometry  is exactly  the correspondence of    Kac\textendash Moody algebras to Yang\textendash Mills geometry.  Because the field  \(\mathcal{D}_{bc}\) acts as a  source of geometric origin in Einstein's equations,  TW gravity gives  insight into the origins of  dark energy\cite{Brensinger:2019mnx}\,,  candidates for dark matter and dark matter portals through the TW Dirac action \cite{Brensinger:2020gcv}\,,  and the origins of the inflaton\cite{Abdullah:2022dzz}\;. Future work is focused on the phenomenology of the fermion coupling, longitudinal modes in gravitational radiation, a projective geometric source for super massive black holes and the interpretation of the $\l$ integration in the context of renormalization group.            
\section{Acknowledgements}
 This work is dedicated to the immense intellectual energy  that Prof.\,Balachandran shares selflessly with all of his collaborators.



\end{document}